# Self-organising management of Grid environments

Ioannis Liabotis[1], Ognjen Prnjat[1], Tope Olukemi[1], Adrian Li Mow Ching[1], Aleksandar Lazarevic[1], Lionel Sacks[1], Mike Fisher[2], Paul McKee[2]

[1]University College London, Dept. of Electronic and Electrical Eng., London WC1E 7JE, UK
email: {iliaboti | oprnjat | tolukemi | aching | a.lazarevic | lsacks}@ee.ucl.ac.uk; Fax: +44 20 7388 9325
[2]BTexact Technologies, email: {mike.fisher | paul.mckee}@bt.com

**Abstract:** This paper presents basic concepts, architectural principles and algorithms for efficient resource and security management in cluster computing environments and the Grid. The work presented in this paper is funded by BTExacT and the EPSRC project SO-GRM (GR/S21939).

## 1 Introduction

Emerging communication and distributed computing models rely on resources being peered with each other, with no centralised management point. Increasingly, these systems are required to be adaptive, self-configuring, autonomous and automated, relieving the users and managers of the need to have detailed knowledge of the system. With highly heterogeneous systems it is almost impossible for one actor to have complete knowledge and control of the environment.

Here we present management architecture aimed at fulfilling the above requirements, for commercial Grid services. Grid computing provides wide-spread dynamic, flexible and coordinated sharing of geographically distributed heterogeneous networked resources, among dynamic user groups. Grid resources are geographically distributed, heterogeneous, and administered within different domains, and their availability is highly dynamic and conditional upon local constraints. In this context the provision of "always-on" services requires a distributed resource management mechanism with capabilities beyond traditional centralised management.

Our management architecture relies on the deployment of light-weight, self-organising and policy-controlled management elements. These elements are seen as an add-on the established Grid platforms such as Globus. The first issue we focus on is resource management and discovery. The process resource consumption has to be monitored and managed so as to guarantee the process QoS as well as maintaining Service Level Agreements (SLAs). Also, processes need to be placed on the optimal node, from the resource availability perspective: thus the need for efficient resource discovery. Second issue we focus on is security management. The nodes of the Grid platform host the user-specified processes: the security requirements are both for access control and intrusion detection. Also, the dynamically changing security information has to be efficiently distributed between the Grid nodes. The basic policy-based resource and security management concepts were developed in [Liab01] [Oluk02] [Prnj02] and their applicability for Grid scenarios was discussed in [Sack03]. In this paper we focus on the general Grid-oriented architecture and specifics of resource discovery and intrusion detection. Section 2 discusses the management principles, the network and functional architecture, and the role of SLAs and policies. Section 3 focuses on the specifics of resource discovery, and section 4 on intrusion detection. Section 5 concludes the paper and section 6 gives the references.

## 2 Management architecture

The overall architecture shown in Figure 1a) focuses on future Grid resource sharing scenarios where resources are shared in collaboration with other parties across multiple domains. Domain A, a commercial Grid operator, has a Service Level Agreement (SLA) with a Grid customer who is allowed to use the resources in A on a commercial basis. Domains A and B also have an SLA regarding resource sharing that might typically exist between the commercial operators. In our architecture, the policy-based management (PBM) [Slom94] principles are used to support the SLAs. SLAs are decomposed into relevant policies that enforce the necessary levels of control at different points in the system. PBM approach thus allows each domain administrator to specify and enforce the behaviour at various levels. It also allows flexibility to be built into the system by supporting a range of behaviours instead of hard-coding a particular one.

SLA between an operator and a user defines (or is decomposed to define) a measurable quantity of resources available to the user, under what conditions, guaranteeing a level of service. The resources allocated to this user are grouped together to form a virtual organisation (VO) subset. Certain higher-level policies will then control the global behaviour of a VO.

Policy-based management in a VO is required to control what resources are being used by the user, to detect if the user exceeds the allocated amount of resources and to prioritise the resources depending on the level of service specified in the SLA. An operator will need to know when to stop allowing resource usage to the user, and to implement the necessary actions for preventing excess usage. Operator might also choose to manage the resources of the VO by running low priority jobs on loaded machines, and use the under-utilised nodes for higher priority jobs. Also, the existence of an SLA between operators would allow overflow if the Grid in one domain is reaching capacity or cannot service a request due to a lack of required resources.

The lower-level policies exert control on the node level by controlling the local node resource management, security / intrusion detection, and resource discovery components (Figure 1b). Discovery service (section 3) is responsible for finding the resources required by a user. Using a self-organised algorithm, it forwards a request to an optimal node capable of performing the task. Security management (section 4) is responsible for managing security and detecting intrusion on the Grid nodes. The local resource management functionality was discussed in the context of active networks in [Prnj02][Oluk02] and in the context of the Grid in [Sack03] and will not be discussed here in detail. Our local management components are to be interfaced with the existing Globus infrastructure: GRAM / Local Job Manager [Czaj98] for enforcing control actions; with Ganglia [Mass03] for monitoring; and with GSI (Globus Security Infrastructure) for key/certificate management. Policies can be distributed using the MDS (Metacomputing Directory Service) [Czaj01] or alternative approaches such as the gossip protocol [Woko02].

Management information regarding resource monitoring, availability, resource usage authorisation and access rights, has to be distributed to the nodes across the domains that are part of the different VOs that are formed. The management information distribution model affects the performance and effectiveness of the management platform. Accurate information capture is vital for resource monitoring and accounting purposes. In a multi-domain Grid scenario we argue two reasons against information distribution on a peer to peer basis. Firstly, excess traffic that floods a network is undesirable; secondly, the information capture of the system at any point in time will be less accurate if information needs to be sent to every other node in the same VO. Our approach is to establish a self-organised overlay network of nodes containing the crucial management information. To achieve this structure all the machines could initially try to gather information from each other using small world [Dunc99] topological network links. Over time we would expect to find a system where the nodes would naturally form a small worlds topology, where a small number of nodes become the regularly used, and queried nodes. The structure of the links should then self-organise as the number of nodes within a virtual organisation changes. This idea is further discussed in the context of resource discovery protocol (section 3).

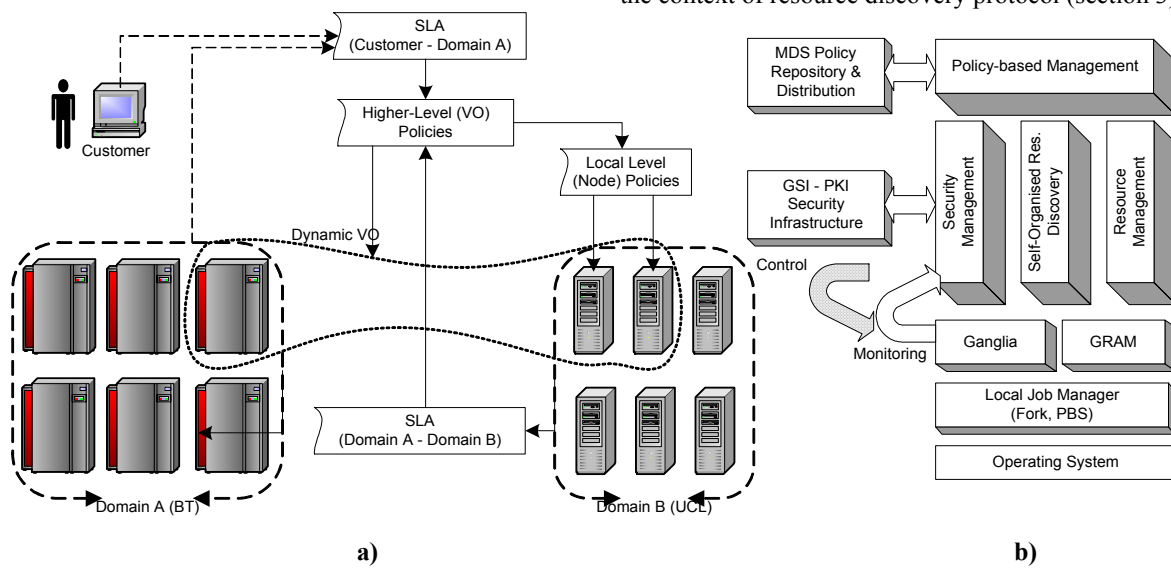

**Figure 1 - Overall architecture (a) and the Node architecture (b)**

# 3 Resource discovery

In a heterogeneous system such as the computational Grid, different types of resources exist and availability is not implied in every node. A resource in this context can be a hardware resource: CPU cycles, encoding hardware, memory/storage or network resources. A resource can also be a specific management interface: special types of resource managers, or job queuing mechanisms, network protocols, and hard resource allocation mechanisms. Finally a resource can be a software service running on a particular node e.g. a web server, operating systems etc. All 3 types of resources appear and disappear dynamically in Grid environment. There are resources that can be considered mostly static, such as operating system versions and special hardware devices. On the other hand resources such as CPU, memory, I/O are much more dynamic.

A basic service of a Grid environment is the resource discovery service. Users request a number of resources with specific properties in order to run their applications. The discovery service is responsible for finding those resources and forwarding the user request to the nodes that can handle it. In an environment where thousands of nodes and hundreds of users are going to use the resource discovery service a centralised mechanism is a potential performance bottleneck and single point of failure for the discovery process [Bal02]. Both in the case of a wide-area Grid and of a computer cluster a decentralised protocol can be used to provide self-organisation. The proposed solution for resource discovery is a fully distributed policy-controlled protocol [Liab03], based on the ideas of peer-to-peer computing [P2P]. The protocol does not guarantee to find the best available resources but it tries to identify an acceptable solution that will be close to the optimum. Each node of the network acts autonomously in order to identify the best possible node that can serve a specific request. Each node is connected to a number of other nodes called neighbours.

The use of a 2-level control mechanism is proposed. The first mechanism, query-based, will be invoked when the node receives a resource request. Associated with each query is a query time to live (QTTL) that determines the number of times a query can be forwarded from a node to its neighbours. The second control mechanism, advertisement-based, is activated when a new node appears and advertises its resources. It is also initiated when a dynamic resource, such as CPU availability, is changing state. Advertisements have their time to live (ATTL) that determines the number of hops that they can be forwarded. Information about the resource availability in the network of nodes is distributed using the query-replies and advertisements. This information is cached in the nodes that receive the replies/advertisements. Nodes that are not included in the future reply messages are removed from the list after a specified time period or when the list exceeds a specified size. Using a different cache for each type of resource, different virtual network topologies are created for each different resource. Initially the nodes are connected to its nearest neighbours (topologically: based on geographical location and bandwidth availability) and some random far neighbours creating a small world topology. As requests are coming to the nodes and the query and advertisement protocols are activated the caches of resource availability are populated with state information. The list of neighbours is changing giving separate lists of neighbours for the different types or resources, thus generating different virtual topologies. Members of the virtual topology are nodes that have the specific resources that the virtual topology identifies or "know" about the requested resources.

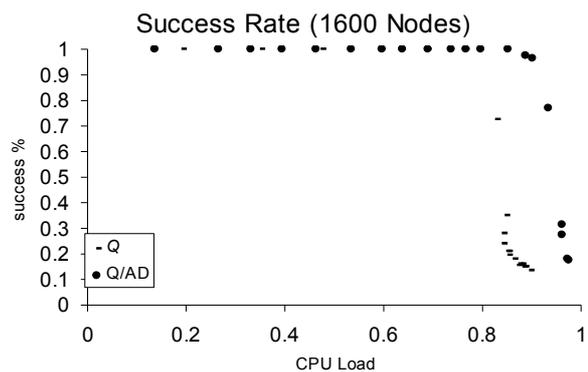

**Figure 2 - Success rate vs. CPU load**

Initial simulations conducted aim to identify the success rate of the protocol for different network loads. To calculate the success rate, the simulation discovers the best node with each new request arrival, using both our resource discovery protocol and a centralised method that obtains each node's load directly. The success rate is the number of times a request was forwarded to the least loaded node over the total number of requests (*i.e.*, the centralised case). Figure 2 shows the relationship between success rate and the mean server load in a simulated Grid network (1600 nodes). Two different variations of the resource discovery protocol where simulated. The first uses only queries; the second uses both queries and advertisements. For mean server load up to around 80% both protocols have 100% success. While the success rate of the query-only version of the protocol drops rapidly when the utilisation is higher than 80%, when advertisements are used the success rate is high until the mean server utilisation is more than 90%. These preliminary results show that our distributed resource discovery protocol provides a high success rate in discovering resources even for high server loads. More studies of this protocol would identify the relationship of the success rate and the management traffic that this protocol generates.

## 4    Security: intrusion detection

Node security management was discussed in the context of active networks in [Prnj02][Oluk02] and in the context of the Grid in [Sack03]. Node security management deals with process and deployer authorisation / authentication. This part of the security management functionality will not be discussed here in detail: the focus is intrusion detection.

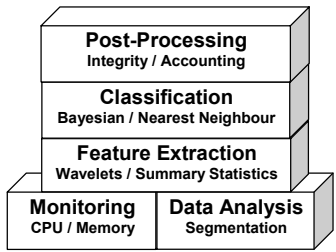

**Figure 3 - Immune agents and the constituent building blocks**

Intrusion detection is performed using the $I^3$ (Integrity, Information and Intelligence) management sub-component. This is an extensible, pluggable XML-based framework, built on concepts from the immune system for pattern recognition and information processing [Dasg97], which aims to provide system integrity and intelligence (for accounting purposes) on distributed open platforms such as the Grid. The system consists of agents operational on each node which cooperate and share information [Stan98]. The objective is to create a totally distributed system with no central control and with little or no hierarchical control which should facilitate the detection of coordinated multi-domain attacks [Cheu97]. The agents on each node should function and interact with other agents based on a simple set of local rules. Each agent is composed of the building blocks shown in Figure 3. The building blocks are code fragments which are fitted together to form a functioning immune agent. At the base the agents monitor processes. The monitored variables are the CPU usage, memory usage and network bandwidth usage over a period of time. The data is analysed and relevant features are extracted from it. Currently the features extracted are the summary statistics of the data. Subsequently the process is classified and based on the classification, post-processing is performed. This could be for intrusion detection or accounting purposes.

The initial post-processing performed with the system has been for intrusion detection. For this post-processing we used a nearest-neighbour classifier [Fuku90]. An agent was initially trained using training data set of normal processes, based on this a threshold was set for a normal process. The features (summary statistics) of the normal process were extracted and along with the threshold value these were passed to a neighbouring node as an XML file (immunisation process). A test data set of normal and abnormal processes was tested on the immunised node; some success was achieved in the immunisation process. The graph in Figure 4 shows the error rate (number of false positives and false negatives) obtained at the immunised node using various threshold values. Figure 4 also shows how we have reduced the error rate by introducing an additional parameter in the immunisation process - a prior probability indicating the probability of process in the test data set being normal. The results here are preliminary results and within the scope of the project we will investigate the use of wavelets for feature extraction and also look into improving the classification method in order to minimise the error rate.

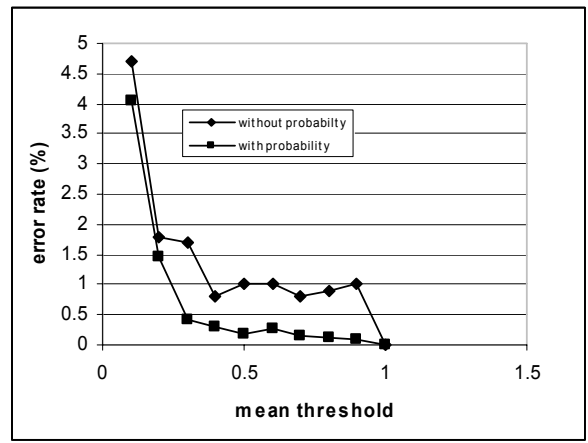

**Figure 4 - Percentage error rate vs. threshold value**

## 5    Conclusion

Management of future Grid environments will require mechanisms which enable efficient resource and security control while maximising automation, flexibility and adaptability and minimising heavy-weight centralised solutions and amount of management traffic. In this paper, we have presented the policy-based management architecture, at whose core are the self-organising light-weight resource management and discovery and security management / intrusion detection components. The proposed algorithms were simulated and the initial results indicate that our approach can be effective in the Grid scenarios.